# Performance Analysis and Prediction in Educational Data Mining: A Research Travelogue

Pooja Thakar
Assistant Professor
VIPS, GGSIPU
Delhi, India

Anil Mehta, Ph.D
Professor
University of Rajasthan
Jaipur, India

Manisha, Ph.D
Associate Professor
Banasthali University
Jaipur, India

## ABSTRACT
In this era of computerization, education has also revamped itself and is not limited to old lecture method. The regular quest is on to find out new ways to make it more effective and efficient for students. Nowadays, lots of data is collected in educational databases, but it remains unutilized. In order to get required benefits from such a big data, powerful tools are required. Data mining is an emerging powerful tool for analysis and prediction. It is successfully applied in the area of fraud detection, advertising, marketing, loan assessment and prediction. But, it is in nascent stage in the field of education. Considerable amount of work is done in this direction, but still there are many untouched areas. Moreover, there is no unified approach among these researches. This paper presents a comprehensive survey, a travelogue (2002-2014) towards educational data mining and its scope in future.

## 1. INTRODUCTION
In last decade, the number of higher education universities/institutions have proliferated manifolds. Large numbers of graduates/post graduates are produced by them every year. Universities/Institutes may follow best of the pedagogies; but still they face the problem of dropout students, low achievers and unemployed students.

Understanding and analyzing the factors for poor performance is a complex and incessant process hidden in past and present information congregated from academic performance and students' behavior. Powerful tools are required to analyze and predict the performance of students scientifically.

Although, universities/institutions collect an enormous amount of students' data, but this data remains unutilized and does not help in any decisions or policy making to improve the performance of students.

If, Universities could identify the factors for low performance earlier and is able to predict students' behavior, this knowledge can help them in taking pro-active actions, so as to improve the performance of such students. It will be a win-win situation for all the stakeholders of universities/institutions i.e. management, teachers, students and parents. Students will be able to identify their weaknesses beforehand and can improve themselves. Teachers will be able to plan their lectures as per the need of students and can provide better guidance to such students. Parents will be reassured of their ward performance in such institutes. Management can bring in better policies and strategies to enhance the performance of these students with additional facilities. Eventually, this will help in producing skillful workforce and hence sustainable growth for the country.

Analysis and prediction with the help of data mining techniques have shown noteworthy results in the area of fraud detection, predicting customer behavior, financial market, loan assessment, bankruptcy prediction, real-estate assessment and intrusion detection. It can be very effective in Education System as well. It is a very powerful tool to reveal hidden patterns and precious knowledge, which otherwise may not be identified and difficult to find and comprehend with the help of statistical methods.

Substantial work is done towards the usage of data mining techniques in Education, but still there are many untouched areas and no unified approach is followed. This paper presents a comprehensive literature review of relevant researches done in last decade from year 2002 to 2014.

This paper is divided into following sections. Second section presents the details of researches done in the area of education in tabular form describing methodologies and findings of each research. Third section summarizes these researches. Fourth section concludes and identifies the areas; where more work is required; hence describes the future scope.

## 2. COMREHENSIVE REVIEW OF LITERATURE
A comprehensive literature review of various significant researches in the area of Educational Data Mining ranging from Year 2002 to 2014 is presented below in a categorized tabular form (Table 1).

These researches can be broadly classified into five areas:

2.1 Survey of papers published in Educational Data Mining.

2.2 Predicting Academic Performance with Pre/Post Enrollment Factors.

2.3 Comparison of Data Mining Techniques in predicting academic performance.

2.4 Correlation among Pre/Post Enrollment Factors and Employability.

2.5 Other areas of Education.

**Table 1**

| Category | Year, Author(s) | Methodology | Key Findings |
|---|---|---|---|
| **Survey of papers published in Educational Data Mining** | 2014, Peña-Ayala, Alejandro | Statistical and Clustering Processes | Identified kinds of educational systems, disciplines, tasks, methods, and algorithms. |
| | 2010, Romero, Cristóbal, and Sebastián Ventura | | Listed tasks in educational area resolved through data mining and future lines. Suggested to develop more unified and collaborative studies. |
| | 2009, Baker, Ryan SJD, and Kalina Yacef | | Identified key features of researches in EDM as discovery with models, emergence of public data, tools. |
| | 2007, Romero, Cristóbal, and Sebastian Ventura | | Presented survey on application of data mining on traditional educational systems. Emphasized on the need of much more specialized work. |
| **Predicting Academic Performance with Pre/Post Enrollment Factors** | 2014, Saranya, S., R. Ayyappan, and N. Kumar | Naive Bayes Algorithm | Graphically represented Institutional Growth Prognosis and Students' Progress Analysis. |
| | 2014, Archer, Elizabeth, Yuraisha Bianca Chetty, and Paul Prinsloo | Experimental Usage of Employee Profiling Software | Experimented the usage of a commercial product generally used for employee profiling in corporate, for higher education environment. |
| | 2014, Hicheur Cairns, Awatef, et al. | Clustering Technique | Professionals' data was analysed during training of a consulting company. |
| | 2014, Arora, Rakesh and Dharmendra Badal | Association Analysis Algorithm | Found set of weak students based on graduation and post graduation marks. |
| | 2012, Osmanbegović, Edin, and Mirza Suljić | Chi-Square Test, One R-Test, Info Gain and Ratio Test, Naive Bayes, DTree | Found predicting model for academic performance that is user friendly for professors or non-expert users. |
| | 2012, Sukanya, M., S. Biruntha, Dr S. Karthik, and T. Kalaikumaran | Bayesian Classification Method | Analysed and assisted the low academic achievers in higher education. |
| | 2011, Torenbeek, M., E. P. W. A. Jansen, and W. H. A. Hofman | Structural Equations Modeling, Correlation Matrix | Examined two variables, Pedagogical approach and skill development in the first 10 weeks of enrollment |
| | 2011, Yongqiang, He, and Zhang Shunli | Association Rules Analysis | Guidance provided for scientific management and comprehensive evaluation of students. |
| | 2011, Sakurai, Yoshitaka, Tsuruta, and Rainer Knauf | Decision Tree | Estimated success chances of curricula by implementing student profiling with storyboard system. |
| | 2011, Aher, Sunita B., and L. M. R. J. Lobo | Classification and Clustering | Analyzed the performance of final year students. |
| | 2010, Ayesha, Shaeela, Tasleem, Ahsan, Inayat | K-Means Clustering | Analyzed students' learning behavior to check the performance of students and predicted weak students. |
| | 2010, Kovacic, Zlatko | Classification Tree Models | Investigated enrolment attributes to pre-identify success of students. |
| | 2010, Al-shargabi, Asma A., and Ali N. Nusari | Clustering, Association Rules and Decision Trees | Analyzed students' academic achievement, students' drop out, and students' financial behavior. |
| | 2010, Yan, Zhi-min, Qing Shen, and Bin Shao | Rough Set Theory | Students' grades were analyzed. |
| | 2010, Ningning, Gao | Neural Network, Rough Set Theory | Predicted drop outs from course |
| | 2010, Knauf, Rainer, Yoshitaka Sakurai, Setsuo Tsuruta, and Kouhei Takada | Decision Tree | Analyzed successful Storyboard (e-learning system) success paths for students. |
| | 2010, Wu, X., Zhang, H., & Zhang, H. | Decision Tree | Suggested comprehensive evaluation method of that can objectively distinguish the grades of students. |
| | 2010, Youping, Bian Xiangjuan Gong | Decision Tree | Evaluated the high school students and studying effectiveness. |
| | 2010, Liu, Zhiwu, and Xiuzhi Zhang | Decision Tree | Built forecasting model for students' marks to identify negative learning habits or behaviors of students. |
| | 2009, Zhu, Li, Yanli Li, and Xiang Li | Association Rule | Predicted student's achievement systematically and improved teaching management. |
| | 2009, Nayak, Amar, Jitendra , Vinod, Shadab | Proposed Use Of Ontology, RDF, XML | Proposed enterprise framework to identify suitable semantic data related to students, faculties and courses. |
| | 2009, Wang, Pei-ji, Lin Shi, Jin-niu Bai, Yu-lin Zhao | Apriori Algorithm | Improved algorithm used to mine the students' data table. |
| | 2009, Ramasubramanian, Iyakutti, and Thangavelu | Rough Set Theory | Predicted weak students. |
| | 2008, Selmoune, Nazih, and | Association Rules, | Found the success and failure factors of students. |
| | Zaia Alimazighi | Apriori Algorithm | |
| | 2008, Shangping, Dai, and Zhang Ping | Genetic Algorithm, Novel Spatial Mining Algorithm | Predicted final grade based on features extracted from log data in web-based system. |
| | 2008, Bresfelean, Paul, Mihaela, Nicolae, Comes | J48 and farthest first algorithms | Provided managerial information on understanding, predicting and preventing academic failure. |
| | 2008, Zhang, Xiaolong, and Guirong Liu | Statistical Approach, Association Rules, | Found hidden patterns for students to avoid becoming low performer ones. |
| | 2008, Villalon, Jorge J., and Rafael A. Calvo | Grammar Trees, Regular Expressions | Concept Map Mining (CMM) done from essays written by students to judge their knowledge. |
| | 2007, Hien, Nguyen Thi Ngoc, and Peter Haddawy | Bayesian Network | Concluded that socio-economic environment can play a major role in the performance of students |
| | 2006, Radaideh, Qasem, Shawakfa, Mustafa | Decision Tree, Rough Set Theory, Naïve Bayes | Produced system generated rules to predict the final grade in a course under study. |
| | 2005, Rasmani, Khairul A., and Qiang Shen | Fuzzy Approaches | Fuzzy approaches used to classify students academic performance |
| | 2005, Delavari, Naeimeh, Reza Beikzadeh, and Somnuk Phon-Amnuaisuk | Decision Tree | Described a roadmap for the application of data mining in higher education by pre- identifying weak students. |
| | 2004, Salazar, A., J. Gosalbez, I. Bosch, R. Miralles, and L. Vergara | Clustering and Decision Rule | Concluded that more variables are required for realistic analysis of academic performance. |
| | 2003, Minaei-Bidgoli, Behrouz, and William F. Punch | Genetic Algorithm | Predicted final grade of student based on features extracted from logged data in an education Web-based system. |
| | 2002, Kotsiantis, S., C. Pierrakeas, and P. Pintelas | Decision Tree, Naive Bayes Algorithm, NN | Concluded that learning algorithms could enable tutors to predict student performance long before final examination. |
| **Comparison of Data Mining Techniques in predicting academic performance** | 2011, Sharma, Mamta, and Monali Mavani | Decision Tree, Sota, Naïve Bayes | Comparison of three algorithms in terms of prediction of students result. |
| | 2009, Siraj, Fadzilah, and Mansour Ali Abdoulha | Cluster Analysis, NN, Logistic Regression and Decision Tree. | Compared three techniques for understanding undergraduate's student Enrolment data. |
| | 2009, Wook, Muslihah, Yuhanim, Norshahriah, Rizal, Isa, Nor Fatimah, and Hoo Yann Seong | Artificial Neural Network, Clustering and Decision Tree | Compared two data mining techniques for predicting and classifying students' academic performance. |
| | 2008, Romero, Cristóbal, Sebastián Ventura, Pedro G. Espejo, and César Hervás | Statistical Classifier, Decision Tree, Rule Induction, Fuzzy, NN | Concluded that classifier model is appropriate for educational use in terms of accuracy and comprehensibility for decision making. |
| | 2007, Nghe, Nguyen Thai, Paul Janecek, and Peter Haddawy | Decision Tree and Bayesian Network | Decision tree proved to be consistently 3-12% more accurate than the Bayesian network in predicting academic performance. |
| **Correlation among Pre/Post Enrollment Factors and Employability** | 2014, Pool, Lorraine Dacre, Pamela Qualter, and Peter J. Sewell | Exploratory and Confirmatory factor analyses, t-test | Emotional Intelligence, Self-Management, Work & Life Experience are found to be important factors for Employability Development Profile. |
| | 2014, Vanhercke, Dorien, Nele De Cuyper, Ellen Peeters, and Hans De Witte | Conceptual | Described that perceived employability is tied to competences and dispositions. |
| | 2013, Potgieter, Ingrid, and Melinde Coetzee | Co-relational Statistics, Multiple Regression | Significant relationships found between the Participants' personality preferences and their employability attributes. |
| | 2013, Finch, David J., Leah K. Hamilton | Two-phase, mixed-methods study | Revealed employers place the highest importance on soft-skills and the lowest importance on academic reputation. |
| | 2013, Bakar, Noor Aieda Abu, Aida Mustapha, and Kamariah Md Nasir | Clustering Analysis Using K-Means and Expectation Maximization Algorithms | Found that across industries, graduates have average interpersonal communication, lacks in creative and critical thinking, analytical and team work. |
| | 2012, Jackson, Denise, and Elaine Chapman | online survey | Significant differences found between academic and employer skill ratings suggesting prominent skill gap between institutes and corporate. |
| | 2011, Gokuladas, V. K | Statistical : Correlation And Multiple Regressions | GPA and proficiency in English language are important predictors of employability and female students are better performers. |
| | 2010, Gokuladas, V. K | Statistical Analysis: Correlation And Regression Analysis | Concluded that Graduates need to possess specific skills beyond general academic education to be employable. |

| **Other areas of Education** | 2013, Jantawan, Bangsuk, and Cheng-Fa Tsai | Bayesian Method and Decision Tree Method | Presented comparison of classification accuracy between two algorithms to evaluate employees' performance. |
|---|---|---|---|
| | 2012, Dejaeger, Karel, et al | Decision Trees, NN, SVM, Logistic Regression | Determined student satisfaction for a particular course. |
| | 2012, Srimani, P. K., and Malini M. Patil | Classifiers, Random Tree, Random Forest | Faculty evaluation based on different parameters |
| | 2012, Yadav, Surjeet Kumar, and Saurabh Pal | Decision Tree | Assisted in selecting Students for enrollment in a particular course. |
| | 2011, Pandey, Umesh Kumar, and Saurabh Pal | Association Rule | Concluded that mix medium class is more preferred over Hindi and English medium class. |
| | 2010, Balakrishnan, Julie M. David | Decision Tree and Clustering | Predicted learning disabilities of school-age children. |
| | 2009, Linjie, Qu, and Lou Lanfang | Modified Apriori Algorithm | Proposed to build evaluation index system and teaching index method based on data mining. |
| | 2009, Ahmed, Almahdi Mohammed, Norita Md Norwawi, and Wan Hussain Wan Ishak | Association Rules | Found the patterns in matching organization and student interests, where they meet each other's requirements. |
| | 2008, Dimokas, Nikolaos, Nikolaos Mittas, Alexandros Nanopoulos, and Lefteris Angelis | Data warehouse | Proposed data warehouse to facilitate and provide thorough analysis of department's data. |
| | 2008, Zhao, Hua-long | OLAP, Data warehouse | Analyzed curriculum's establishment from many angles. |
| | 2008, Pumpuang, Pathom, Anongnart Srivihok, and Prasong Praneetpolgrang | Bayesian Network, Decision Forest | Nbtree identified as the best classifiers to predict student sequences for course registration planning |
| | 2008, Ranjan, Jayanthi, and Saani Khalil | Conceptual | Described data mining process in management education and focusing on academic aspects of admission and counseling process. |
| | 2007, Bresfelean, Vasile Paul | Decision Trees | Analyzed students' choice in continuing their education with post University studies (Master degree, PhD) |
| | 2006, Aksenova, Svetlana S., Du Zhang, and Meiliu Lu | Support Vector Machines and Rule-Based Approach | Proposed an approach to predict the enrollment headcount by a predictive model built from new, continued and returned students. |
| | 2004, Talavera, Luis, and Elena Gaudioso | Naive Bayes | Described the usage of Data Mining Techniques to support evaluation of collaborative activities. |

Noteworthy research papers and their findings are mentioned below in each category.

## 2.1 Survey of papers published in Educational Data Mining

Recent paper published in 2014 in Elsevier titled "Educational Data Mining: A Survey and a Data Mining-Based Analysis of Recent Works" presented the survey of published papers from 2010-2013 and divided Educational Data Mining approaches in kinds of educational systems, disciplines, tasks, methods, and algorithms. Author identified that each Educational Data Mining approaches can be organized according to six functionalities student modeling, student behavior modeling, assessment; student performance modeling; student support and feedback versus curriculum-domain knowledge-sequencing, mostly focusing on academic performance [6].

Romero and Ventura, in 2010 published a paper in IEEE, which listed most common tasks in the educational environment resolved through data mining and some of the most promising future lines. Educational Data Mining community remained focused in North America, Western Europe, and Australia/New Zealand. They mentioned that there is a considerable scope for an increase in educational data mining's scientific Influence. They also suggested developing more unified and collaborative studies [45]. In another paper by them in year 2007 titled "Educational data mining: A survey from 1995 to 2005" surveyed the application of data mining to traditional educational systems. They concluded that much more specialized work is needed in order for educational data mining to become a mature area.[69]

Paper titled "An Empirical Study of the Applications of Data Mining Techniques in Higher Education" published in year 2011, listed potential areas in which data mining can be applied in higher education [31].

## 2.2 Predicting academic performance with Pre/Post Enrollment Factors

Most of the published research papers belong to this category. Latest work published in International Journal of Computer Science and Mobile Computing, 2014 describes the process of finding the set of weak students based on graduation and post graduation marks [5]. Another paper published in European Journal of Scientific Research in 2010 also analyzed students' learning behavior to predict weak students. [34]. P. Ramasubramanian, K. Iyakutti and P. Thangavelu, in year 2009 also predicted weak students using rough set theory. [54]

To assist in selecting students for enrollment in a particular course Surjeet Kumar Yadav and Saurabh Pal used Decision Trees technique of data mining in 2012 [22]. In another paper in 2010 Zlatko J. Kovačić also investigated enrolment attributes to pre-identify success of students.[35]

In 2012, M. Sukanya, S. Biruntha, Dr.S. Karthik and T. Kalaikumaran analyzed and assisted the low academic achievers in higher education using Bayesian Classification Method of Data Mining. [21]

A comprehensive evaluation method for undergraduates; that can objectively distinguish the grades of students was developed by Xiewu, Huacheng Zhang, Huimin Zhang in year 2010 [40]. Another study by Dai Shangping, Zhang Ping, in year 2008 predicted final grades of students based on features extracted from log data in web-based system and published their work in IEEE [56]

Paper published in IEEE in 2011 titled "Success Chances Estimation of University Curricula Based on Educational History, Self-Estimated Intellectual Traits and Vocational Ambitions" integrated student profiling with storyboard system (Learning Management System) and concluded that success chance heavily depends on individual properties.[26]

Jorge J. Villalon and Rafael A. Calvo did a novel study named as Concept Map Mining (CMM) from essays written by students to judge their knowledge and published their work in IEEE, 2008 [61]

## 2.3 Comparison of Data Mining Techniques in predicting academic performance of students

In 2011, Springer published a paper by Mamta Sharma and Monali Mavani for comparison of three algorithms in terms of prediction of students result [32].

In 2009 Fadzilah Siraj and Mansour Ali Abdoulha compared three techniques for understanding undergraduate's student enrolment data and published their work in IEEE [47]. Nbtree was identified as the best classifiers to predict student sequences for course registration planning in paper published by Pathom Pumpuang, Anongnart Srivihok and Prasong Praneetpolgrang in year 2008 in IEEE [62].

Cristóbal Romero, Sebastián Ventura, Pedro G. Espejo and César Hervás concluded that classifier model is appropriate for educational use in terms of accuracy and comprehensibility for decision making in 2008 [64].

Decision tree proved to be consistently 3-12% more accurate than the Bayesian network in predicting academic performance of undergraduate and postgraduate students in a paper titled "A Comparative Analysis of Techniques for Predicting Academic Performance" in 2007 [68], IEEE.

## 2.4 Correlation among Pre/Post Enrollment Factors and Employability

Recent work published in Emerald Group Publishing Limited, 2014 clearly stated that emotional intelligence, self-management, life experiences are important factors for Employability Development Profile (EDP) [1]. Another work published in Emerald Group Publishing Limited, 2014 described that employability is in strong correlation with competences and dispositions [7].

In 2014, Cairns, Gueni, Fhima, David and Khelifa analyzed employees' profiles during training of a consulting company involved in training of professionals for employability skills and found positive correlation in their jobs/assignments, history etc. [3].

In SA Journal of Industrial Psychology, 2014, authors, Potgieter & Coetzee, observed a number of significant relationships between the participants' personality and employability. Set of eight core employability attributes identified to increase the likelihood of securing and sustaining employment opportunities [8].

David J. Finch, Leah K. Hamilton, Riley Baldwin and Mark Zehner in year 2013 did 30 one-on-one interviews with hiring managers and 115 employers and revealed employers place the highest importance on soft-skills and the lowest importance on academic reputation [12]. In another online survey by Denise Jackson and Elaine Chapman in 2012 reported significant differences between academic and employer skill ratings suggesting prominent skill gap between institutes and corporate [13].

Two papers published by Wiley, authored by V. K. Gokuladas in year 2010 and 2011 respectively. In first paper he reflected that graduates need to possess specific skills beyond general academic education to be employable [33]. In next paper he showcased that GPA and proficiency in English language are important predictors of employability [24].

In 2013, Noor Aieda, Abu Bakar, Aida Mustapha, Kamariah Md. Nasir collected primary data from The Ministry Of Higher Education, Malaysia and found that across industries, graduates have average interpersonal communication skills, lacks in creative and critical thinking, problem solving, analytical skills , and team work [10]. In 2013, Bangsuk Jantawan and Cheng-Fa Tsai, made an effort to design a model for supporting the prediction of the employees' performance in an organization using data mining techniques [9].

## 2.5 Other areas of Education

In 2012, International Conference on Intelligent Computational Systems published a paper titled "A Classification Model For Edu-Mining" for faculty evaluation based on different parameters [20]. Another paper published in 2009 in IEEE proposed to build evaluation index system and teaching index method based on data mining [49].

International Journal of Computer Science Issues 2011, published a paper titled "A Data Mining View On Class Room Teaching Language", which concluded that mix medium class is more preferred over Hindi and English medium class [27].

Hua-Long Zhao, in his paper titled "Application of OLAP to the Analysis of the Curriculum Chosen by Students" published in IEEE in 2008, analyzed curriculum's establishment from many angles [60]. Prediction of learning disabilities of school-age children was done by Julie M. David and Kannan Balakrishnan in 2010 [43]. In 2007, Vasile Paul Brefelean analyzed students' choice in continuing their education with post University studies (Master degree, PhD) using data mining techniques [67].

Svetlana S. Aksenova, Du Zhang, and Meiliu Lu proposed an approach to predict the enrollment headcount by a predictive model built from new, continued and returned students and published their work in IEEE in 2006 [71]. In 2004, Luis Talavera and Elena Gaudioso described the usage of Data Mining Techniques to support evaluation of collaborative activities [74]. Two papers published by IEEE in 2010 predicted students' drop outs [36] [38].

# 3. SUMMARY

In summary, all of these researches done previously, reveals some significant areas in education field, where prediction with data mining has reaped benefits; such as finding set of weak students [5], determining student's satisfaction for a particular course [14][50], Faculty Evaluation [20], Comprehensive student evaluation [25][26][40], Class room teaching language selection [27], Predicting students' dropout [34] [36], course registration planning [62], predicting the enrollment headcount [71], evaluation of collaborative activities [74] etc.

Few researches have shown significant relationships between the participants' personality preferences and their employability attributes [8]. It is observed that there are specific skills that graduates need to possess in order to become employable and that these skills are beyond general academic education. [33]. Attributes like emotional intelligence, self-management, and Work & life experience are also important factors for Employability Development Profile (EDP) [1]. Employers also place the highest importance on soft-skills and the lowest importance on academic reputation [12]. Thus, perceived employability is tied to competences and dispositions, rather than only on academic qualification [7]. Significant differences are found between academic and employer skill ratings suggesting prominent skill gap between academia and corporate [13].

# 4. FUTURE PROSPECTS

One of the most recent and biggest challenge that higher education faces today is making students skillfully employable. Many universities/institutes are not in position to guide their students because of lack of information and assistance from their teaching-learning systems. To better administer and serve student population, the universities/institutions need better assessment, analysis, and prediction tools.

Considerable amount of work is done in analyzing and predicting academic performance, but all of these works are segregated. There is a clear need for unified approach. Other than academic attributes, there are large numbers of factors that play significant role in prediction, which includes non-cognitive factors (set of behaviors, skills, attitudes). Suitable data mining techniques are required to measure, monitor and infer these factors for prediction. Thus enriching the input vector with qualitative values may increase the accuracy rate of prediction as well.

Integrated Models/Frameworks are required for all the stakeholders of an Institution; hence ensuring sustainable growth for all (Management, Teachers, Students and Parents).